\documentclass[openany]{nature}

\usepackage[T1]{fontenc}
\usepackage[left]{lineno}
\usepackage{amsthm,amsmath,amssymb}
\usepackage{mathrsfs}
\usepackage{overpic}
\usepackage{multirow}
\usepackage[T1]{fontenc}
\usepackage{graphicx}
\usepackage{pdflscape}
\usepackage{hyperref}
\usepackage{threeparttable}
\usepackage{threeparttablex}
\usepackage{xcolor}
\usepackage{ulem}
\usepackage{cancel}
\usepackage[figuresleft]{rotating}
\usepackage{caption}
\usepackage{subfigure}
\usepackage{tablefootnote}
\usepackage{hyperref}
\usepackage{graphicx}
\usepackage{longtable}
\usepackage{supertabular}
\usepackage{float} 
\usepackage{authblk}
\usepackage{multibib}
\usepackage{subfiles}
\usepackage{enumerate}
\usepackage{bm}
\usepackage{booktabs}

\hypersetup{
colorlinks=true,
linkcolor=red,
filecolor=blue,
urlcolor=cyan,
citecolor=green}

\makeatletter
 
\def\emph#1 {\textit{ #1 } }

\let\saved@includegraphics\includegraphics
\AtBeginDocument{\let\includegraphics\saved@includegraphics}
\renewenvironment*{figure}{\@float{figure}}{\end@float}
\makeatother

\newcommand{\araa}{Annu. Rev. Astron. Astrophys.}  
\newcommand{\aap}{Astron. Astrophys.}  
\newcommand{\mnras}{Mon. Not. R. Astron. Soc.}  
\newcommand{\prl}{Phys. Rev. Lett.}  
\newcommand{\sci}{Science} 

\definecolor{dkblue}{RGB}{54, 86, 169}

\def\be{\begin{eqnarray}}
\def\ee{\end{eqnarray}}

\makeatletter

\title{A high-energy neutrino flare associated with nearby bright interacting supernova SN 2021foa}

\author{Ming-Xuan Lu$^{1,2}$, Yun-Feng Liang$^{1,2}$\thanks{E-mail: liangyf@gxu.edu.cn}, Xue-Rui Ouyang$^{1,2}$, Da-Bin Lin$^{1,2}$, Xiang-Gao Wang$^{1,2}$\thanks{E-mail: wangxg@gxu.edu.cn}, Yi-Zhong Fan$^{3,4}$}

\begin{document}
\maketitle

\begin{affiliations}

\item Guangxi Key Laboratory for Relativistic Astrophysics,
School of Physical Science and Technology, Guangxi University, Nanning 530004, China
\item GXU-NAOC Center for Astrophysics and Space Sciences, Nanning 530004, People's Republic of China
\item Key Laboratory of Dark Matter and Space Astronomy, Purple Mountain Observatory, Chinese Academy of Sciences, Nanjing 210023, China
\item School of Astronomy and Space Science, University of Science and Technology of China, Hefei 230026, China

\end{affiliations}

\begin{abstract}
While core-collapse supernovae have been widely discussed as potential neutrino sources, definitive observational evidence has remained elusive. In this work, we report evidence of high-energy neutrino emission in the direction of supernova SN 2021foa, which is one of the closest and brightest interacting supernovae observed to date. Using the second data release of muon track data from the IceCube Neutrino Observatory, we conducted a time-dependent analysis and identified a neutrino clustering that temporally coincides with the optical peak of SN 2021foa, occurring approximately 16 to 22 days after the discovery date, with a maximum test statistic of $\sim 28.2$. 
Monte Carlo simulations indicate that the probability of observing such a neutrino excess by chance is $\sim6.7 \times 10^{-5}$, corresponding to a significance of $\sim4.0\,\sigma$. The spatial and temporal correlation strongly suggests that the neutrinos originate from the supernova. SN 2021foa is a unique "flip-flop" supernova; its spectra repeatedly transitioned between hydrogen-rich (Type IIn) and helium-rich (Type Ibn) phases within 50 days post-peak, reflecting a violent and complex mass-loss history of its progenitor. 
The inferred neutrino energy exceeds the optical radiative energy and ejecta kinetic energy of the supernova by orders of magnitude, suggesting that the neutrino emission is likely powered by a delayed central engine driving a jet that is choked within the dense circumstellar medium.
\end{abstract}

The discovery of high-energy astrophysical neutrinos \cite{IceCube:2013low,IceCube:2014stg} has opened a new era of neutrino astronomy. The diffuse astrophysical neutrino flux detected by the IceCube Neutrino Observatory indicates the existence of neutrino sources in the universe. Through multi-messenger observations, several neutrino sources have been identified or supported by strong evidence, including the blazar TXS 0506+056 \cite{IceCube:2018dnn,IceCube:2018cha}, the Seyfert galaxy NGC 1068 \cite{icecube2022evidence}, and the Galactic plane \cite{icecube2023observation}. Additionally, several transient sources, such as the tidal disruption events (TDEs) AT2019dsg, AT2019fdr, and AT2019aalc, have been proposed as potential neutrino sources \cite{2021NatAs...5..510S,2022PhRvL.128v1101R,2024MNRAS.529.2559V,Li:2024qcp}. However, these sources collectively account for only a small fraction of the observed diffuse neutrino flux, suggesting the presence of other significant populations of neutrino sources \cite{Murase:2015xka,Murase:2016gly,Sridhar:2022uis}.

Core-collapse supernovae (CCSNe) have long been considered promising candidate sources of high-energy neutrinos \cite{Murase:2010cu,Murase:2017pfe,Senno:2015tsn,Zhu:2021mqc}. Shock collisions between the supernova ejecta and the dense circumstellar medium (CSM) can accelerate cosmic-ray protons, which subsequently interact with protons (proton-proton, or $pp$, interactions) or photons (photohadronic, or $p\gamma$, interactions) in the CSM to produce high-energy neutrinos. These two mechanisms are expected to yield distinct spectral signatures: the $pp$ scenario typically produces a softer neutrino spectrum ($\gamma \gtrsim 3$) due to strong proton cooling effects, whereas the $p\gamma$ scenario tends to produce a harder spectrum ($\gamma \sim 2$). Furthermore, relativistic jets launched during core collapse that fail to penetrate the progenitor envelope or the external CSM (referred to as "choked jets") can also serve as efficient sites for neutrino production \cite{Meszaros:2001ms,Razzaque:2004yv,Ando:2005xi,PhysRevLett.111.121102}. However, despite these theoretical expectations, dedicated IceCube searches for supernovae have yet to find any statistically significant neutrino excess \cite{IceCube:2021oiv,IceCube:2023esf,Chang:2022hqj,IceCube:2023ogt}, and individual candidates (such as the recently reported SN 2017hcd \cite{Ji:2026vnq}) remain to be fully confirmed. Some candidate associations have also been claimed by cross-matching neutrino alert events with astronomical catalogs \cite{IceCube:2015jsn,Lu:2025jks,Stein:2025uxi}, however the statistical significance of these single-event associations is usually limited, hard to provide compelling evidence for high-energy neutrino emission from supernovae.

Type IIn supernovae (SNe IIn) are characterized by the presence of narrow Balmer emission lines in their spectra, indicating ongoing interactions between the ejecta and a dense, hydrogen-rich CSM \cite{1990MNRAS.244..269S,1997ARA&A..35..309F}. Due to their exceptionally high CSM densities, SNe IIn are considered particularly promising sites for neutrino production \cite{Murase:2010cu}. SN 2021foa is a peculiar Type IIn supernova, notable for repeatedly transitioning its spectral type (switching from hydrogen-rich Type IIn to helium-rich Type Ibn, and then back to Type IIn) within approximately 50 days post-peak, earning it the designation of a "flip-flop" supernova \cite{Farias:2024ilk}. This behavior has been attributed to helium-rich ejecta sweeping through a dense, hydrogen-rich shell post-explosion \cite{Farias:2024ilk}, or alternatively, to a complex CSM consisting of two layers with distinct density structures \cite{Gangopadhyay:2024pik}. Its progenitor is hypothesized to be a massive star transitioning from a luminous blue variable (LBV) to a Wolf-Rayet (WR) phase \cite{Gangopadhyay:2024pik,2022A&A...662L..10R}, having undergone violent mass-loss episodes in the years to months prior to the explosion. SN 2021foa reached a peak absolute magnitude of $M_r \approx -18$ mag in the barred spiral galaxy IC 863 ($z = 0.0084$, $d \approx 34.9$ Mpc). Its light curve exhibits a double-peaked structure characteristic of SN 2009ip-like events, consisting of a precursor outburst followed by the main peak \cite{2022A&A...662L..10R}. Light-curve modeling indicates that the progenitor experienced extreme pre-explosion mass loss at rates of up to $\sim 2\ M_{\odot}\ \text{yr}^{-1}$ \cite{Farias:2024ilk,Gangopadhyay:2024pik}, nearly three orders of magnitude higher than typical Type II supernovae, which created an exceptionally dense CSM around the star. Compared to all SNe IIn cataloged in the Zwicky Transient Facility Bright Transient Survey (ZTF BTS) \cite{Perley:2020ajb}, SN 2021foa is both the optically brightest (peak $r \approx 15.4$ mag) and the closest (Extended Data Fig.~\ref{efig2}), making it a prime target for searching for high-energy neutrino emission from Type IIn supernovae with current observational capabilities.

Given that SN 2021foa exhibits both the dense CSM environment characteristic of Type IIn and the helium-rich ejecta features of Type Ibn, and represents the closest and optically brightest Type IIn supernova in the current sample, we select it as a high-priority target for a time-dependent high-energy neutrino search. In this paper, we report evidence of high-energy neutrino emission in the direction of SN 2021foa. 

\section*{Results}
\subsection{Time-dependent likelihood analysis}

We conduct a time-dependent neutrino source search at the position of SN 2021foa using the second data release (DR2) of muon track data from the IC86 detector configuration of the IceCube Neutrino Observatory \cite{Abbasi:2026ehs}. Our analysis employs an unbinned maximum likelihood method \cite{Braun:2008bg}, where the likelihood function incorporates spatial, energy, and temporal probability density function (PDF) terms. The signal spatial PDF is defined by convolving the angular reconstruction error of each event with a Gaussian centered at the source location. The signal energy PDF is modeled assuming a power-law neutrino flux, $\Phi(E) \propto E^{-\gamma}$, convolved with the IceCube effective area and detector smearing matrix. For the signal temporal PDF, we adopt a Gaussian time profile, representing a neutrino flare distributed symmetrically around a central time $T_0$. The background PDFs are constructed directly from the observed data distribution. Details regarding the likelihood formalism are provided in the Methods section; the likelihood analysis is implemented using the official {\tt SkyLLH} software package.

The free parameters in our analysis are the neutrino spectral index $\gamma$, the number of signal events $n_s$, the flare central time $T_0$, and the Gaussian width $\sigma$. The likelihood function is maximized with respect to these parameters to evaluate the test statistic, defined as $TS = 2 \log(\mathcal{L}_{\text{max}} / \mathcal{L}_{\text{null}})$. To ensure fitting stability, a lower bound of $\sigma \ge 1$ day is imposed. The spectral index $\gamma$ is allowed to vary freely between 1.0 and 10.0, and the search window for $T_0$ was constrained to $\pm 100$ days around the supernova discovery date (MJD 59288.45).

\begin{figure}
\centering
\includegraphics[width=0.99\textwidth]{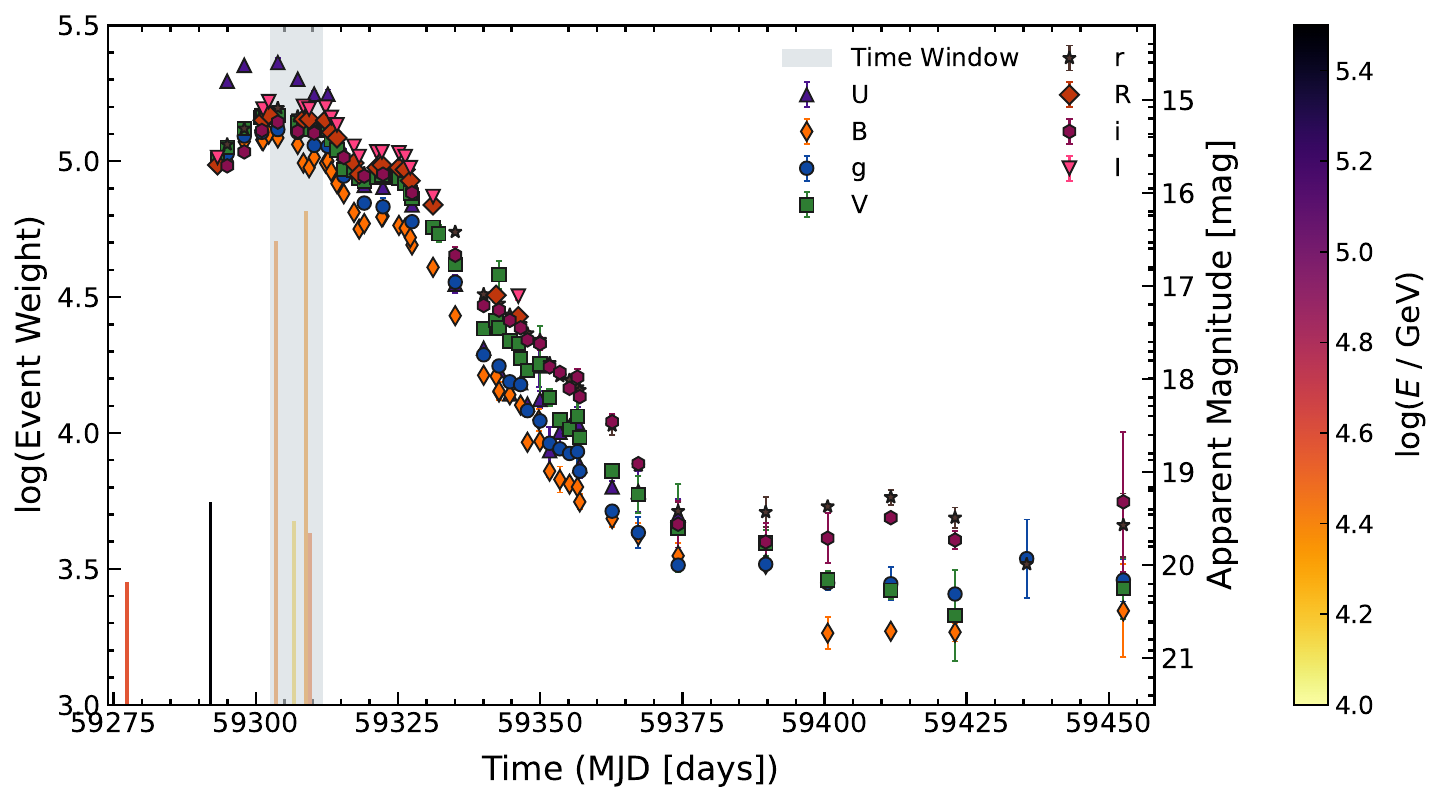}
\caption{\noindent\textbf{Optical light curves and neutrino flare of SN~2021foa.}
Multi-band optical apparent magnitudes of SN~2021foa \cite{Gangopadhyay:2024pik} (right-hand axis; symbol shapes and colours correspond to photometric filters as indicated in the legend) are shown together with the logarithm of the signal event weights from the time-dependent neutrino analysis (left-hand axis). The vertical bars mark the arrival times of individual neutrino events, with their colours encoding the reconstructed muon energy, $\log(E/\mathrm{GeV})$, according to the colour scale. The grey shaded region denotes the best-fit Gaussian time window of the neutrino flare ($T_0\pm2\sigma$ with $T_0 = \mathrm{MJD}\,59307.1$ and $\sigma \approx 2.3$\,days). The four clustered neutrino events ($\sim$15--30~TeV) arrived within this window, temporally coincident with the optical maximum.}
\label{fig1}
\end{figure}

We identified a high-$\rm TS$ neutrino emission excess near the optical peak of SN 2021foa (see Fig.~\ref{fig1}). The best-fit parameters are as follows: central flare time of $T_0 = \text{MJD}\ 59307.1$ (approximately 19 days after the optical discovery date, coinciding closely with the $r$-band peak time of $\text{MJD}\ 59302.35$); Gaussian flare width of $\sigma \approx 2.3$ days; spectral index of $\gamma = 4.9 \pm 1.6$; signal event number of $n_s = 3.86 \pm 2.0$; test statistic of $\rm TS = 24.7$; and the muon neutrino flux normalization at 1 TeV of $\sim4.2 \times 10^{-9}\ \text{GeV}^{-1}\ \text{cm}^{-2}\ \text{s}^{-1}$.  This neutrino excess is primarily driven by four events, all located within $\sim 1^\circ$ of SN 2021foa (with angular distances ranging from $0.16^\circ$ to $0.86^\circ$). Their arrival times span a window of approximately 6 days, with reconstructed muon energies in the range of $\sim 15$–$30$ TeV. This remarkably soft spectrum ($\gamma \approx 4.9$) resembles, but is even softer than, the neutrino spectrum observed for NGC 1068 ($\gamma \approx 3.1$–$3.5$) \cite{icecube2022evidence}. Such a soft spectral index could suggest strong cooling effects in the neutrino production region, though it may also be a result of statistical fluctuations considering the large uncertainty on the spectral index.

\begin{figure}
\centering
\includegraphics[width=0.8\textwidth]{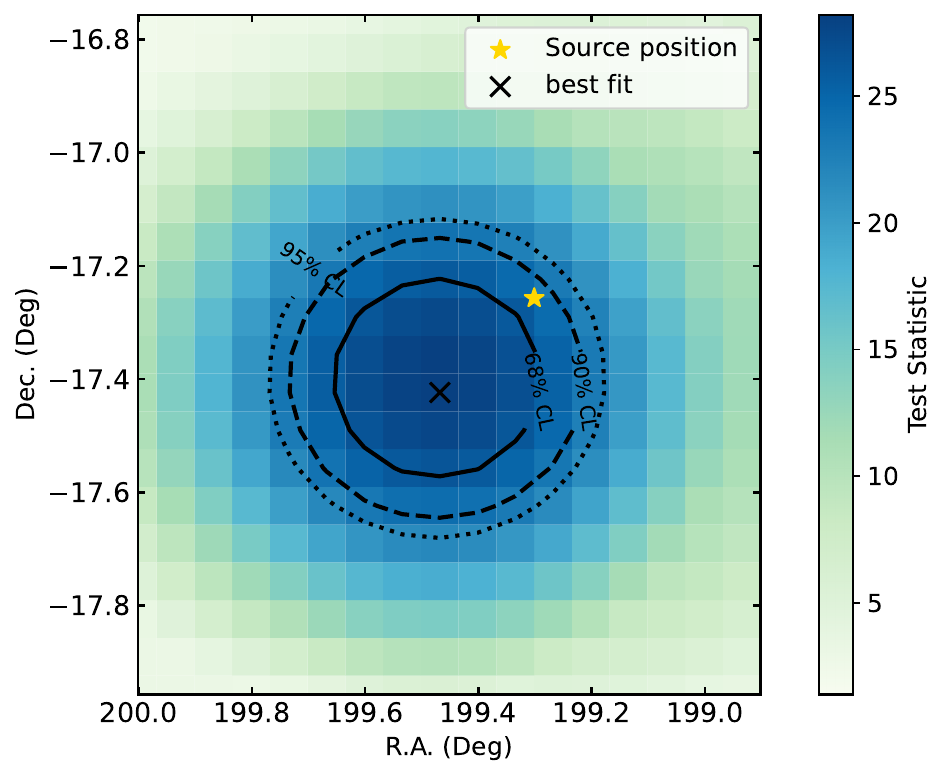}
\caption{\textbf{Test-statistic map from the neutrino analysis around SN~2021foa.}
The map shows the TS values evaluated on a grid of celestial coordinates around the optical position of SN~2021foa. The colour scale encodes the TS value. The yellow star marks the known optical position of SN~2021foa, and the black cross indicates the best-fit neutrino source position from the maximum likelihood analysis. The solid, dashed, and dotted contours denote the 68\%, 90\%, and 95\% confidence-level error regions, respectively. The best-fit position is consistent with the optical location of the supernova within the 90\% confidence level, supporting a physical association between the neutrino excess and SN~2021foa.}
\label{fig2}
\end{figure}

To further examine the spatial association of the neutrino excess with SN~2021foa, we constructed a TS map by evaluating the test statistic on a grid of celestial  coordinates around the optical position of the supernova (Fig.~\ref{fig2}). The best-fit neutrino source location is consistent with the optical location of SN~2021foa within the 90\% confidence-level error region. This spatial consistency reinforces the physical association between the neutrino signal and the supernova.
At the best-fit neutrino source position, the likelihood analysis gives a maximum TS of $\sim 28.2$ for the neutrino excess.

\subsection{Significance estimation}

To evaluate the global significance of the observed signal, we perform Monte Carlo simulations. Following the identical analysis pipeline used for SN 2021foa, we randomly selected coordinate points in the southern sky (declination $\delta < -10^\circ$) and random central times within the 11-year IC86 data. We then executed the same neutrino flare search within a $\pm 100$-day window around each selected time. Out of 15,000 trials, only one yielded a $TS$ value exceeding the observed value of 24.7 (Extended Data Fig.~\ref{efig1}). This corresponds to a $p$-value of $\sim6.7 \times 10^{-5}$, equivalent to a significance of $\sim4.0\sigma$.

Given that the signal associated with SN 2021foa is primarily driven by four neutrino events clustered within $\sim 1^\circ$ of the source, we specifically evaluate the probability of such spatial and temporal clustering arising purely from background fluctuations. We randomly selected coordinate points in the southern sky ($\delta < -10^\circ$) and random central times $t_0$ across the IC86-I through IC86-XI operating periods. We then recorded the number of events falling within a $1^\circ$ radius of the coordinates during a 6-day window starting from $t_0$ (i.e., $[t_0, t_0 + 6\ \text{days}]$). Across 100,000 trials, we found no instances of $\ge 4$ events clustering within the selected regions (with the maximum observed being 3 events). This corresponds to a $p$-value of $<1 \times 10^{-5}$.  This low $p$-value indicates that producing a spatio-temporal clustering of this magnitude purely through background fluctuations is highly improbable. Crucially, this rare clustering coincided with the peak of the optical light curve of SN 2021foa—the most energetic phase of the supernova explosion. The statistical rarity of this spatial clustering ($p < 1 \times 10^{-5}$), combined with its tight temporal alignment with the optical peak, provides a compelling, self-consistent argument suggesting that this neutrino excess is highly likely to have originated from SN 2021foa itself.

\subsection{Isotropic Neutrino Energy}

Based on the best-fit parameters of $n_s = 3.86 \pm 2.0$ and $\gamma = 4.9 \pm 1.6$, and adopting a distance of $d = 34.9$ Mpc to SN 2021foa, we estimated the all-flavor isotropic equivalent neutrino energy emitted during the 6-day flare duration (integrated over the energy range of 20 TeV to $10^6$ TeV) to be $E_{\nu, \rm iso} \approx 4.7 \times 10^{52} \ \rm erg$, with a $1\sigma$ uncertainty range spanning from $\sim 5.1 \times 10^{51}\ \rm erg$ to $\sim1.2 \times 10^{53}\ \rm erg$ (see however below for more discussion on the isotropic energy). The wide uncertainty is primarily driven by the large uncertainty in the spectral index $\gamma = 4.9 \pm 1.6$. 
Furthermore, because the spectral parameters are constrained by only four neutrino events, a strong degeneracy exists between $n_s$ and $\gamma$, which further amplifies the uncertainty in the estimated energy.

\section*{Discussion}

Our findings provide strong observational evidence that core-collapse supernovae can produce high-energy neutrinos. The arrival window of the signal (approximately 16 to 22 days after the optical discovery, centered at $\text{MJD}\ 59307.1$) aligns closely with the optical peak of SN 2021foa. Considered independently, the probability of four background neutrinos clustering within a $1^\circ$ region over 6 days is less than $1 \times 10^{-5}$. Moreover, that this clustering coincided with the optical peak, i.e., the most energetic stage of the supernova, makes a pure background fluctuation explanation increasingly unlikely. 

The nominal value of the best-fit spectral index ($\gamma \sim 4.9$) is significantly softer than the value of $\sim2.0$ predicted for proton spectra under diffusive shock acceleration (DSA), where the neutrino spectrum roughly inherits the spectral index of the parent protons in $pp$ interactions. The spectral index is also softer than most previously reported neutrino sources. However, such a soft spectral index is physically plausible. Because SN 2021foa is located in the southern sky, the dominant atmospheric muon background necessitates stringent energy cuts for southern events. This causes the IceCube effective area in this direction to drop to zero below $\sim 20$ TeV \cite{Abbasi:2026ehs}, meaning the detector only probes the high-energy tail of the spectrum. Consequently, the observed soft index could correspond to an exponential cutoff part of the neutrino spectrum, which may be attributed to the maximum particle acceleration energy ($E_{\rm max}$) being severely limited by $pp$ collision cooling, proton radiative cooling, or the cooling of secondary pions and muons \cite{Murase:2022dog}.
Notably, the neutrino spectrum of NGC 1068 ($\gamma \approx 3.1$–$3.5$) is similarly interpreted as reflecting a highly obscured, strongly cooled environment near its core \cite{Murase:2019vdl}; SN 2021foa could represent an extreme manifestation of this phenomenon. Furthermore, given the large statistical uncertainty on the fitted spectral index, a harder spectrum similar to that of NGC 1068 cannot be ruled out (e.g., $\gamma \sim 3.3$–$3.5$ remains well within the $1\sigma$ uncertainty interval).

The aforementioned isotropic energy of $4.7 \times 10^{52}\,\text{erg}$ (all-flavor, with a $1\sigma$ uncertainty range from $5.1 \times 10^{51}\,\text{erg}$ to $1.2 \times 10^{53}\,\text{erg}$) is based on the neutrino flux directly detected within IceCube’s observable energy band.  Because SN 2021foa is located in the southern sky ($\delta \approx -17^\circ$), the IceCube effective area for muon tracks in this region cuts off below $E_\nu\sim 20$ TeV, and the four detected events (with reconstructed muon energies of $E_{\rm rec}\sim 15$–$32$ TeV) lie precisely on the rising edge of this effective area. Consequently, the detected events are highly likely to represent only the high-energy tail of a much larger, low-energy neutrino population. Roughly assuming the actual $E_{\nu,\rm iso}$ is an order of magnitude higher than the current estimate of $4.7 \times 10^{52}\,\text{erg}$, reaching $\sim 5 \times 10^{53}\,\text{erg}$, this energy is much higher than the total optical radiative energy ($\sim 10^{50}\,\text{erg}$ \cite{Farias:2024ilk}) and is also about two orders of magnitude higher than the ejecta kinetic energy of SN 2021foa ($E_{\rm kin} \approx 3.0 \times 10^{51}\,\text{erg}$ \cite{Gangopadhyay:2024pik}). This indicates that the neutrino emission is unlikely to be driven solely by the interaction between the ejecta shock and the CSM, and an additional energy source independent of the ejecta kinetic energy must be introduced.

A possible physical scenario is a delayed central engine driving a choked jet. During the initial $\sim15$ days after core collapse, the optical emission is dominated by the ejecta shock breakout through the dense CSM, corresponding to the optical peak. Subsequently, a central engine (possibly a black hole fallback accretion or a rapidly rotating magnetar) is established at approximately 15 to 20 days and begins injecting energy to drive a jet. Given the extremely dense CSM surrounding SN 2021foa (with a mass-loss rate of up to $\dot{M} \sim 2\,M_\odot\ \text{yr}^{-1}$ \cite{Farias:2024ilk}), the jet is completely choked during its propagation through the CSM, forming a relativistic shock cocoon at the jet head. The ejecta also has sufficient column density to decelerate or choke the jet within the system. Cosmic-ray protons accelerated within the shock cocoon interact with target protons in the surrounding dense CSM via $pp$ collisions, producing high-energy neutrinos through $\pi^\pm \to \nu$ cascades. Since the shock cocoon is surrounded by a dense medium, the energy of cosmic-ray protons is efficiently converted (even approaching the calorimetric limit of $pp$ interactions) into pions and subsequently decays into neutrinos.

In the calorimetric limit, the pion production efficiency is $f_\pi \approx 1$, while the energy fraction carried by all-flavor neutrinos in the pion decay chain is $f_{\pi\to\nu} \approx 3/8$. Therefore, the required total cosmic-ray energy and the true neutrino energy (after correcting for the geometric effect of jet beaming) satisfy $E_{\rm CR} \approx (8/3)\,E_{\nu,\rm true}$. Assuming a jet half-opening angle of $\theta_j$, the beaming factor is $f_{\rm beam} \approx \theta_j^2/2$, and the true neutrino energy is $E_{\nu,\rm true} = E_{\nu,\rm iso} \cdot f_{\rm beam}$. For a nominal isotropic-equivalent energy of $5 \times 10^{53}\,\text{erg}$, when $\theta_j \approx 2^\circ$, $E_{\nu,\rm true} \approx 3 \times 10^{50}\,\text{erg}$, corresponding to $E_{\rm CR} \approx 8 \times 10^{50}\,\text{erg}$; when $\theta_j \approx 5^\circ$, $E_{\nu,\rm true} \approx 1.8 \times 10^{51}\,\text{erg}$, corresponding to $E_{\rm CR} \approx 4.8 \times 10^{51}\,\text{erg}$. These energies are well within the typical range that a central engine can provide.

The time delay of the neutrino emission ($\sim19$ days after the discovery date) can be interpreted in the choked jet model as the onset timescale of the central engine (such as the formation of a black hole accretion disk and the accumulation of fallback material). This delay mechanism temporally coincides with the shoulder of the optical light curve at $+17$ days \cite{Gangopadhyay:2024pik}, which might reflect the modulation of the fallback accretion process by the geometry of the CSM disk. However, it is more likely that the optical and neutrino emissions are physically independent (neutrinos originating from the engine, whereas the optical emission arises from the ejecta), and their temporal coincidence is purely due to a consistency of timescale orders. Furthermore, the non-detection of SN 2021foa in the radio band \cite{Gangopadhyay:2024pik} is consistent with expectations of a choked jet scenario, where the jet is completely absorbed and no relativistic outflow escapes; meanwhile, the steep index of the neutrino spectrum, $\gamma \approx 4.9\pm1.6$, reflects the typical characteristic of particle acceleration being suppressed or truncated in a dense environment.

Finally, considering that large uncertainties remain in the measurement of $E_{\nu,\rm iso}$ (including both statistical uncertainties and the uncertainty from extrapolation to below $20\,\text{TeV}$), if we adopt the $1\sigma$ lower limit of the statistical uncertainty ($\sim5 \times 10^{52}\,\text{erg}$ after accounting for the extrapolation), the required cosmic-ray energy would only be $\sim 10^{50}$–$10^{51}\,\text{erg}$ even for a relatively wide jet opening angle, making the energy budget much more relaxed. Therefore, within the current observational uncertainties, the delayed central engine plus choked jet model can reasonably explain the high-energy neutrino emission of SN 2021foa.

\captionsetup[table]{name={\bf Table}}
\captionsetup[figure]{name={\bf Fig.}}

\clearpage

\clearpage

\clearpage

\clearpage
\section*{Methods}
\subsection{IceCube Data} 

The IceCube Neutrino Observatory is a cubic-kilometer-scale detector embedded in the Antarctic ice, consisting of 5,160 digital optical modules (DOMs) deployed on 86 strings \cite{IceCube:2016zyt}. This analysis utilizes the recently released DR2 muon-track data from IceCube \cite{Abbasi:2026ehs}, which spans 14 years of observations from April 2008 to May 2022. In this work, we only use data acquired under the complete 86-string detector configuration (IC86-I through IC86-XI, from May 2011 to May 2022, totaling 11 seasons with a cumulative live time of approximately 10.7 years).
Each event in the DR2 dataset includes reconstructed parameters such as arrival time, direction (right ascension and declination), angular uncertainty of the directional reconstruction ($\sigma_i$), and an energy proxy (reconstructed muon energy $E_\mu$). 

The DR2 release also provides binned instrument response functions (IRFs), 
comprising effective areas $A_{\rm eff}(E_\nu, \delta_\nu)$ and five-dimensional 
response matrices (smearing matrices) that encode the mapping from the true neutrino energy 
$E_\nu$ and declination $\delta_\nu$ to the reconstructed muon energy $E_\mu$, 
point-spread function, and angular uncertainty~\cite{Abbasi:2026ehs}. 
In our likelihood analysis, the signal energy and spatial probability density 
functions are constructed by convolving the assumed neutrino spectrum with 
the effective area and the smearing matrix of the IRF.

With a declination of approximately $-17^\circ$, SN 2021foa is located in the southern sky. In this region, the background in the dataset is composed of atmospheric muons (dominating the lower-energy regime) and atmospheric neutrinos (consisting of both conventional and prompt components, the latter arising from charmed meson decays), with a minor contribution from the diffuse astrophysical neutrino flux. Because the atmospheric muon background in the southern sky is orders of magnitude higher than in the northern sky (where atmospheric muons are shielded by the Earth), to suppress the intense atmospheric muon background, the DR2 event selection imposes stringent cuts on southern-sky events~\cite{Abbasi:2026ehs}. As a consequence, the IceCube southern-sky data have a high low-energy threshold and the effective area drops rapidly to zero below $\sim 20$~TeV.

\subsection{Maximum-Likelihood Analysis}

We adopt the {\tt SkyLLH} package~\cite{Abbasi:2026ehs} to implement the standard time-dependent unbinned maximum likelihood analysis \cite{Braun:2008bg,2010APh....33..175B}. 
Here we briefly introduce the formalism of the likelihood analysis. 
The likelihood function is formulated as a mixture of signal and background components:
\begin{equation}
    \mathcal{L}(n_{\mathrm{s}}, \gamma, T_0, \sigma) = \prod_{i=1}^{N} 
    \left[ \frac{n_{\mathrm{s}}}{N} S_i + \left(1 - \frac{n_{\mathrm{s}}}{N}\right) B_i \right],
    \label{eq:like}
\end{equation}
where $n_{\mathrm{s}}$ is the number of signal neutrino events and $N$ is the total number of events in the ROI. The signal and background probability density functions (PDFs) for the $i$-th event are factorized into independent spatial, energy, and temporal terms~\cite{Braun:2008bg,2010APh....33..175B}:
\begin{equation}
    S_i = S^{\rm spat}_i(\vec{x}_i, \sigma_i \mid \vec{x}_s) \cdot 
           S^{\rm ener}_i(E_i \mid \delta_s, \gamma) \cdot 
           S^{\rm temp}_i(t_i \mid T_0, \sigma), \label{eq:sig} 
\end{equation}
\begin{equation}
    B_i = B^{\rm spat}_i(\delta_i) \cdot 
        B^{\rm ener}_i(E_i \mid \delta_i) \cdot 
        B^{\rm temp}_i(t_i). \label{eq:bg}
\end{equation}
The signal spatial PDF $S^{\rm spat}_i$ is modeled as a two-dimensional Gaussian distribution, with its width determined by the angular reconstruction uncertainty  $\sigma_i$ of each event, centered at the source direction $\vec{x}_s$. The signal energy PDF $S^{\rm ener}_i$ is constructed by convolving an assumed power-law neutrino spectrum $dN_\nu/dE_\nu \propto E_\nu^{-\gamma}$ with the IceCube effective area $A_{\rm eff}(E_\nu, \delta_s)$ and the energy-smearing matrix of the IRFs provided in the DR2 data~\cite{Abbasi:2026ehs}. For the signal temporal PDF, a Gaussian profile is adopted, representing a neutrino flare centered at $T_0$ with width $\sigma$:
\begin{equation}
    S^{\rm temp}_i(t_i \mid T_0, \sigma) = \frac{1}{\sqrt{2\pi}\,\sigma} 
    \exp\left[-\frac{(t_i - T_0)^2}{2\sigma^2}\right].
    \label{eq:stime}
\end{equation}
The background spatial and energy PDFs are constructed directly from the observed data distribution within each declination band, following the standard data-driven approach. The background temporal PDF is uniform over the total livetime $T_{\mathrm{L}}$ of the data sample, i.e., $B^{\rm temp}_i = 1 / T_{\mathrm{L}}$.
The expected number of signal events is related to the differential neutrino flux $\Phi(E_\nu) = \Phi_0 (E_\nu / E_0)^{-\gamma}$ by
\begin{equation}
    n_{\mathrm{s}} = \Delta T \int A_{\rm eff}(E_\nu, \delta_s) \, 
    \Phi(E_\nu) \, dE_\nu,
    \label{eq:ns}
\end{equation}
where $\Delta T$ is the duration of the signal and $A_{\rm eff}$ is the effective area at the declination $\delta_s$, and the temporal modulation is accounted for within the likelihood via $S^{\rm temp}_i$.

The likelihood is maximized over four free parameters: the number of signal events $n_{\mathrm{s}} \ge 0$, the spectral index $1.0 \le \gamma \le 10.0$, the flare central time $T_0$ within $\pm 100$~days of the discovery date (MJD~59288.45), and the Gaussian width $\sigma \ge 1$~day (imposed to ensure fitting stability). The test statistic (TS) is defined as the log-likelihood ratio between the best-fit signal hypothesis and the background-only null hypothesis:
\begin{equation}
    \mathrm{TS} = 2 \log\left[ \frac{\mathcal{L}(\hat{n}_{\mathrm{s}}, 
    \hat{\gamma}, \hat{T}_0, \hat{\sigma})}{\mathcal{L}(n_{\mathrm{s}} = 0)} \right],
    \label{eq:ts}
\end{equation}
where the hat notation denotes the best-fit values.

\subsection{Monte Carlo Simulations}
To estimate the statistical significance of the observed $TS$ value, we performed 15,000 background simulations using the 11-year IceCube IC86 dataset (IC86-I through IC86-XI). In each simulation, we randomly selected a coordinate in the southern sky (declination $\delta < -10^\circ$) and a random time within the 11-year observation period. We then performed the same time-dependent search analysis as for SN 2021foa within a $\pm 100$-day window centered on the selected time. The maximum $TS$ value obtained from each search is recorded to construct the null hypothesis distribution of the test statistic. 
In each trial, the source position is fixed at the randomly selected coordinates without any positional scan; we therefore compare the resulting TS distribution against the value of 24.7 obtained at the known optical location of SN~2021foa, rather than the higher value of $\sim 28.2$ that includes an additional optimization over the source position.
The observed $\rm TS$ of 24.7 for SN 2021foa lies at the extreme tail of this distribution. Out of the 15,000 completed trials, only one reached or exceeded this value (the maximum simulated $TS$ is 25.7), which corresponds to a $p$-value of $\sim6.7 \times 10^{-5}$.

To further evaluate the statistical rarity of having four neutrino events cluster within a $1^\circ$ region over a 6-day window, we performed another type of Monte Carlo simulations. In each simulation, we randomly selected a coordinate in the southern sky ($\delta < -10^\circ$) and a random time $t_0$, and then recorded the number of events located within a $1^\circ$ angular distance of the coordinate during the interval $[t_0, t_0 + 6\ \text{days}]$. A total of 100,000 simulations were performed. The maximum number of clustered events observed in any of these trials is 3, and no trial yields $\ge 4$ events.
Therefore, the chance probability of observing such an event clustering in the southern sky is $<1\times10^{-5}$.

\clearpage

\section*{Data Availability}

The IceCube DR2 muon-track data used in this analysis are publicly available at the IceCube Data Archive
(\url{https://icecube.wisc.edu/science/data}) and are described in detail in Ref.~\cite{Abbasi:2026ehs}.
The optical photometric data of SN~2021foa are obtained from the published literature cited in this work.
The processed data presented in the figures of this paper are available from the corresponding author
(Y.-F. L.) upon reasonable request.

\section*{Code Availability}
Upon reasonable requests, the code (mostly in Python) used to produce the results and figures will be provided.

\clearpage


\clearpage

\begin{addendum}

\item[Acknowledgments] 
We acknowledge the IceCube Collaboration for the release of 14-year muon-track data and for providing the \texttt{SkyLLH} software package for neutrino point-source analysis. This work is supported by the National Key Research and Development Program of China (Grant 2022YFF0503304 to Y.-F. L.) and the National Natural Science Foundation of China (Grants 12373042, U1938201, and 12494573 to X.-G. W.). X.-G. W. acknowledges the support from the Bagui Scholars Programme in Guangxi.

\item[Competing Interests] The authors declare no competing interests.

\item[Additional information] Correspondence and requests for materials should be addressed to Yun-Feng Liang.

\end{addendum}

\clearpage
\setcounter{figure}{0}
\setcounter{table}{0}
\captionsetup[table]{name={\bf Extended Data Table}}
\captionsetup[figure]{name={\bf Extended Data Fig.}}

\clearpage
\begin{figure}
\centering
\includegraphics[width=0.85\textwidth]{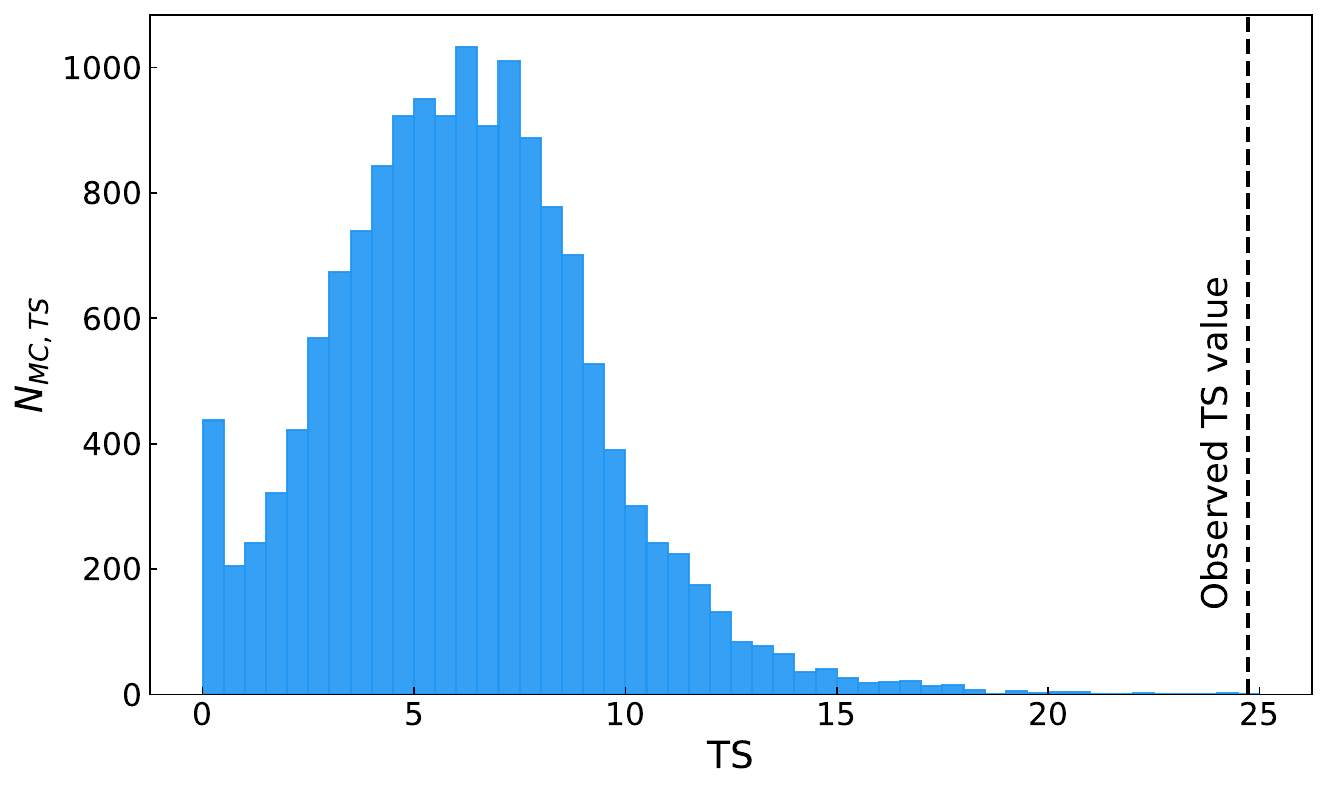}
\caption{\textbf{Distribution of maximum TS values from background Monte Carlo simulations.}
The histogram shows the distribution of the maximum TS values obtained from 15{,}000 background trials, in which random coordinates in the southern sky ($\delta < -10^\circ$) and random central times were searched using the identical time-dependent analysis pipeline as for SN~2021foa. The vertical dashed line marks the observed TS value for SN~2021foa. Only one out of the 15{,}000 trials exceeds this value, corresponding to a $p$-value of $\sim6.7\times 10^{-5}$.}
\label{efig1}
\end{figure}

\clearpage
\begin{figure}
\centering
\includegraphics[width=0.8\textwidth]{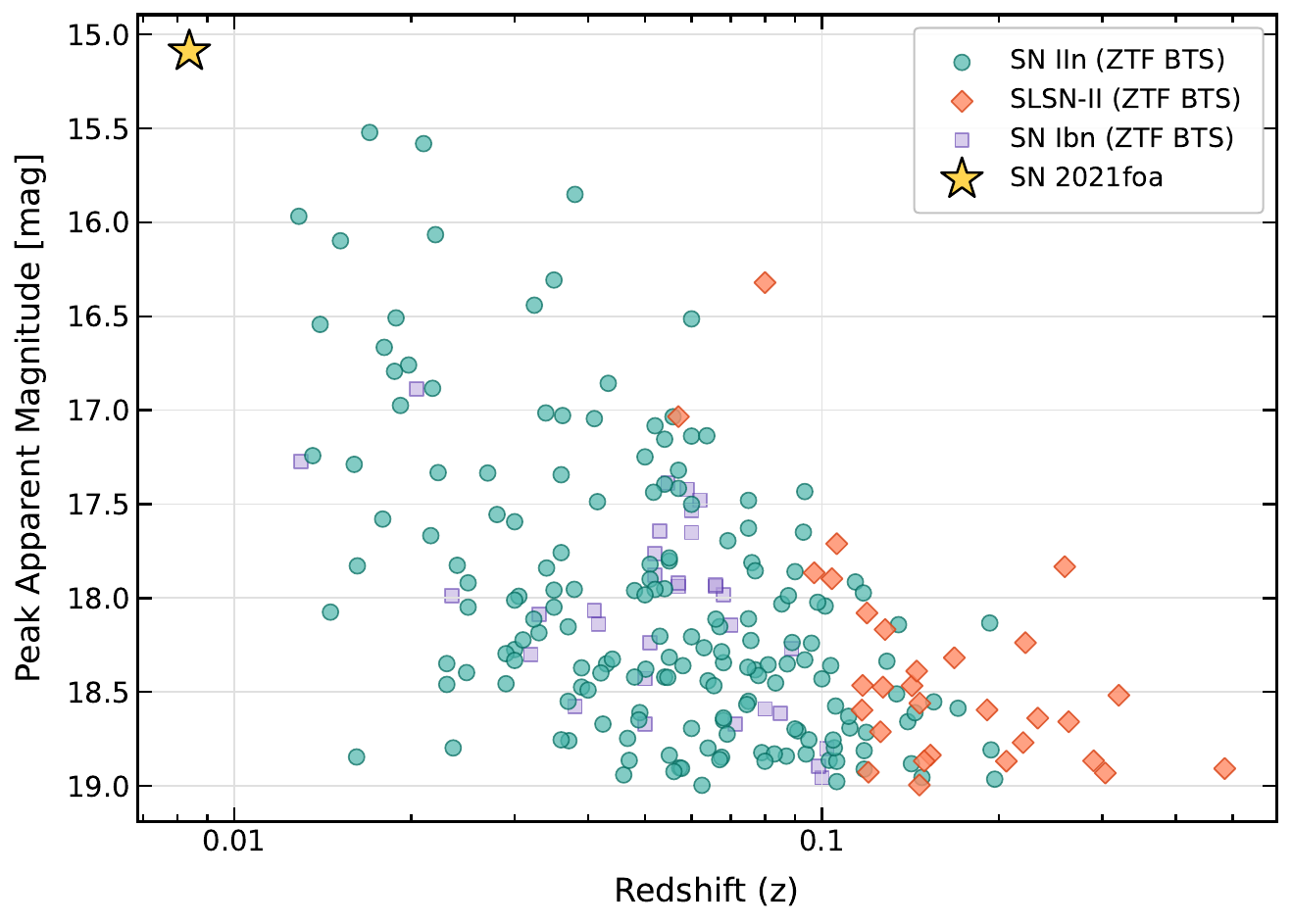}
\caption{\noindent\textbf{Peak apparent magnitude versus redshift for SN~2021foa and other supernovae.} 
SN~2021foa (yellow star) is compared to samples of Type~IIn supernovae (green circles), superluminous supernovae Type~II (SLSN-II; orange diamonds), and Type~Ibn supernovae (purple squares) from the Zwicky Transient Facility Bright Transient Survey (ZTF BTS). SN~2021foa, at $z=0.0084$, is the nearest and optically brightest Type~IIn supernovae compared to the ZTF BTS supernovae, reaching a peak apparent magnitude significantly brighter than the bulk of the comparison population.}
\label{efig2}
\end{figure}

\clearpage
\begin{figure}
\centering
\includegraphics[width=0.8\textwidth]{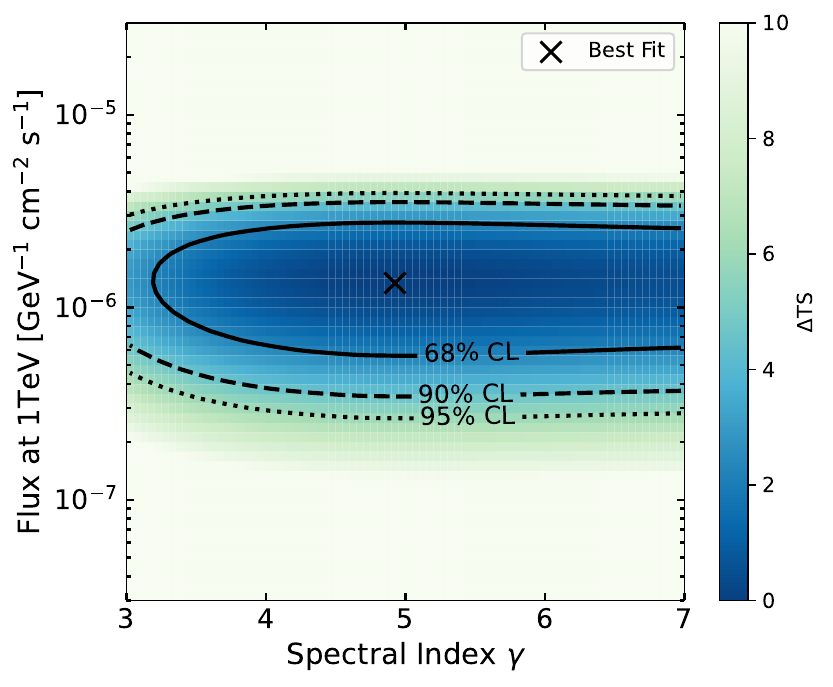}
\caption{\noindent\textbf{Confidence regions for the neutrino spectral index and flux normalization.}
The colour scale shows the difference in TS ($\Delta \mathrm{TS}$) relative to the best-fit point (black cross) as a function of the neutrino spectral index $\gamma$ and the muon-neutrino flux normalization at 1~TeV. The solid, dashed, and dotted contours enclose the 68\%, 90\%, and 95\% confidence-level regions, respectively. The broad elongation of the contours reflects that the spectral index $\gamma$ is poorly determined by the fit due to the small number of detected events.}
\label{efig3}
\end{figure}

\end{document}